\def\year{2017}\relax
\documentclass[letterpaper]{article} 
\usepackage{ICWSM17}  
\usepackage{times}  
\usepackage{helvet}  
\usepackage{courier}  
\usepackage{url}  
\usepackage{graphicx}  
\usepackage{bm}  
\usepackage[utf8]{inputenc}
\usepackage{fourier} 
\usepackage{array}
\usepackage{makecell}
\usepackage{dblfloatfix}
\begin{document}
%
\title{Detecting Socio-Economic Impact of Cultural Investment Through\\ Geo-Social Network Analysis}

\author{Xiao Zhou\\
Computer Laboratory\\
Univerisity of Cambridge\\
xz331@cam.ac.uk
\And
Desislava Hristova\\
Computer Laboratory\\
Univerisity of Cambridge\\
desislava.hristova@cl.cam.ac.uk
\And
Anastasios Noulas\\
Data Science Institute\\
University of Lancaster\\
a.noulas@lancaster.ac.uk
\And
Cecilia Mascolo\\
Computer Laboratory\\
Univerisity of Cambridge\\
cecilia.mascolo@cl.cam.ac.uk
}

\maketitle
\begin{abstract}
Taking advantage of nearly 4 million transition records for three years in London from a popular location-based social network service, Foursquare, we study how to track the impact and measure the effectiveness of cultural investment in small urban areas. We reveal the underlying relationships between socio-economic status, local cultural expenditure, and network features extracted from user mobility trajectories. This research presents how geo-social and mobile services more generally can be used as a proxy to track local changes as government financial effort is put in developing urban areas, and thus gives evidence and suggestions for further policy-making and investment optimization. 
\end{abstract}

\section{Introduction}
In 1997, the striking “Bilbao miracle” created by Guggenheim Museum not only provided Bilbao, a depressed Spanish port town, with a dramatic socio-economic growth, but also provided evidence that city can blossom with cultural investment (González 2011). Even though the ability of cultural investment to promote local regeneration has been widely accepted, large-scale evaluation and prediction of its impact are still not practiced. The potential of network science in providing insight on deprivation dynamics (Eagle et al. 2010) along with the millions of human mobility traces made available by location-based applications has so far been untapped in culture-led regeneration studies. In this paper, we present how geo-social network data from location-based social network Foursquare\footnote{https://foursquare.com} can be used to quantify the effect of cultural investment on the urban regeneration process in London’s neighbourhoods. We define new metrics on cultural investment and cultural features
in geo-social networks to measure the priority level of culture for urban areas and show how differences in these reflect on the network properties of local areas. In addition, we propose an innovative approach to uncovering the underlying relationships between socio-economic status, cultural investment and geo-social network properties using a fusion of network analysis and statistical analysis.

\section{Related Work}
Conventionally, the investigation of socio-economic deprivation for urban areas has largely relied on government statistics, by which data are generally obtained from traditional surveys. It is usually costly to implement and takes a few years to carry out each time. With an aim to overcome this limitation, researchers have recently started to mine low-cost, real-time, and fine-grained new data sources for socio-economic deprivation study. For instance, Eagle et al (2010) discovered a high correlation between call network diversity and urban area deprivation using call records data; Quercia et al (2012) found the topic of tweets and deprivation level of urban area are correlated; Smith et al (2013) using Oyster Card data to identify areas of high deprivation level in London; Quercia and Saez (2014) explored the relationship between the presence of certain Foursquare venues with social deprivation; Venerandi et al. (2015) using Foursquare and OpenStreetMap datasets to explore the correlation between urban elements and deprivation of neighbourhoods; Hristova et al. (2016) discussed the relationship between the prosperity of people and urban places, and distinguished between different categories and urban geographies using Foursquare and Twitter data. Inspired by previous qualitative works in the urban planning domain that found correlations between cultural investment and urban development, as well as quantitative studies from data scientists, we propose to use geo-social network data to detect the change of socio-economic deprivation related to cultural investment in urban environment.

\section{Hypotheses}
As a base condition, we expect that different socioeconomic conditions and amounts of cultural expenditure in neighbourhoods will lead to significantly different network properties in the geo-social graph. Therefore, we hypothesize that:

\textbf{[H1]} Areas with high cultural investment and deprivation level \textit{have significantly different network and local properties} from areas with low cultural investment and deprivation level.

This assertion lays the foundation of further investigation into the nature of culture-led urban regeneration, where based on existing case studies from literature (González 2011), we expect that cultural investment in more deprived neighbourhoods results in growth. Specifically:

\textbf{[H2]} Areas with high cultural investment and deprivation level \textit{experience significant growth} with respect to network and local properties from areas with lower cultural investment and deprivation level.

\section{Datase}

\subsection{Socio-economic Data}

The dataset used to evaluate the socio-economic status is the English Indices of Deprivation, a measure of deprivation for small areas in England calculated by the Department for Communities and Local Government. It produces the Index of Multiple Deprivation (IMD), which reflects the overall deprivation level of areas. The versions used in this research are the 2010 Index\footnote{https://www.gov.uk/government/statistics/english-indices-of-deprivation-2010} and 2015 Index\footnote{https://www.gov.uk/government/statistics/english-indices-of-deprivation-2015}, through
a comparative analysis of which, an insight of socioeconomic deprivation changes of urban areas can be derived.

\subsection{Cultural Expenditure Data}
The cultural expenditure data used in this work is the local authority revenue expenditure and financing\footnote{https://www.gov.uk/government/collections/local-authority-revenueexpenditure-and-financing}. This dataset is based on returns from all 444 local authorities in England, showing how they allocate their yearly spending on various services, including “cultural and related services”. In this study, revenue spending data for the financial years 2010/11, 2011/12, and 2012/13 are involved.

\subsection{Foursquare Data}
Alongside the two official datasets from government introduced above, we also use user mobility records and information of venues in London from Foursquare. It contains all “transitions” (pairs of check-ins by users between two different venues) occurring within London for three years from January 2011 to December 2013. In total, there are 3,992,664 transitions generated between 17804 venues in London during the study period.

\subsection{Geo-social Network}
The Foursquare dataset can be depicted as a spatial network of venues connected by transition flows of users. It is a directed graph where nodes represent start and end venues, while edges correspond to transitions. The weight of the edge is proportional to the number of transitions made by all users between the two venues. We assume the impact of expenditure can be observed after 9 months through Foursquare. Formally, we define our yearly dataset as a directed graph $G_t =(V_t, E_t)$ for $t=1,2,3$, where $t$ indicates the different snapshots in time of the dataset. The network properties are shown in Table 1.

\begin{table}[t]
\small
	\centering
	\begin{tabular}{*{6}{c}}
		\hline
		\textbf{t} & \textbf{Duration} & $\bm{|V|}$ & $\bm{|E|}$ & $\bm{\left \langle C \right \rangle}$ & $\bm{\left \langle k \right \rangle}$ \\
		\hline
		\textbf{1} & 2011.1-2011.12 & 15832 & 469229 & 0.221 & 59\\
		\hline
		\textbf{2} & 2012.1-2012.12 & 16189 & 715113 & 0.228 & 70\\
		\hline
		\textbf{3} & 2013.1-2013.12 & 17684 & 742017 & 0.240 & 84\\
		\hline
	\end{tabular}
	\caption{Network Properties at Each Snapshot: Number of Nodes $|V|$, Number of Edges $|E|$, Average Clustering Coefficient $\left \langle C \right \rangle$, and Average Degree $\left \langle k \right \rangle$.}
\end{table}

\begin{figure}
	\centering
	\includegraphics{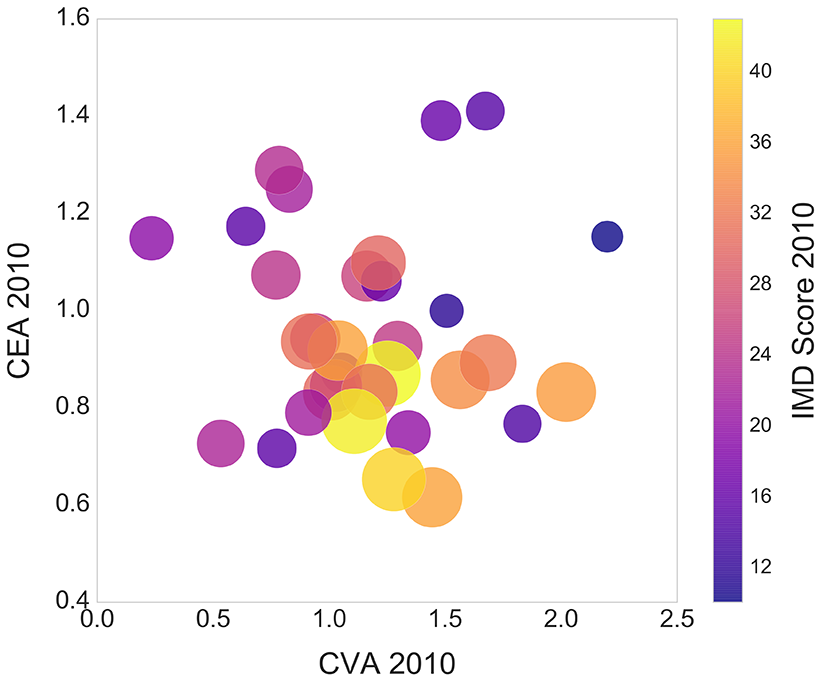}
	\caption{Initial IMD Score, CEA, and CVA of London Boroughs}
\end{figure}

\begin{figure*}[b!]
	\centering
	\includegraphics{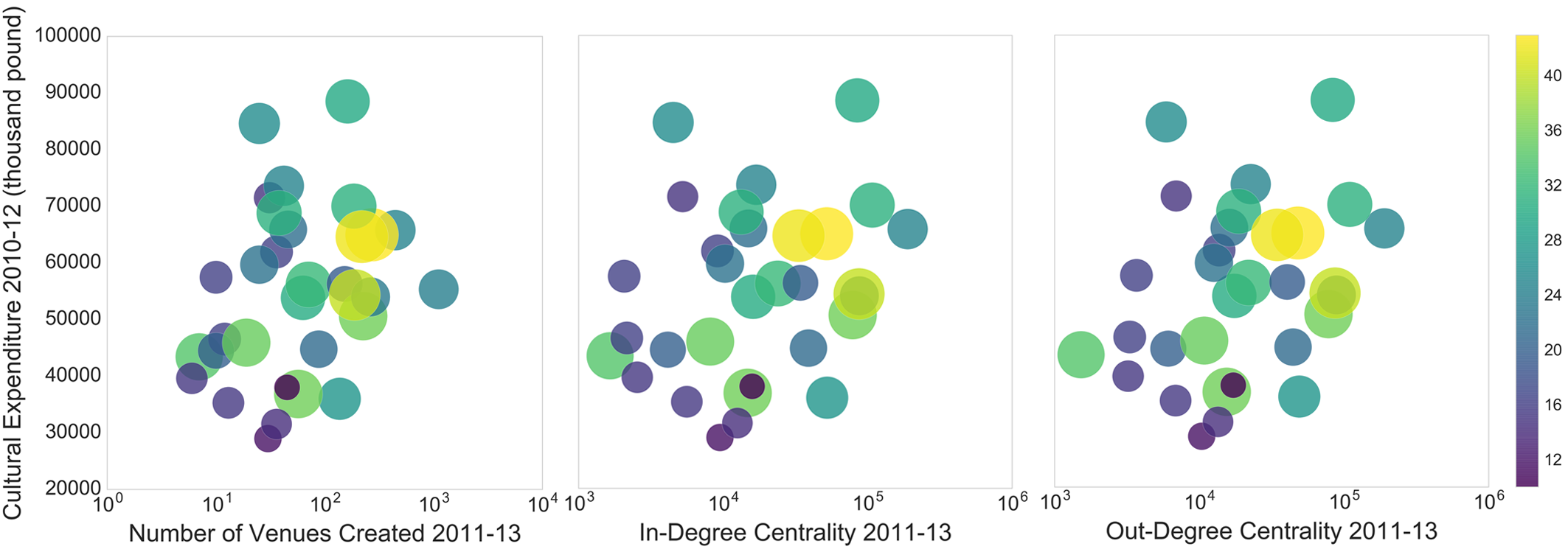}
	\caption{Culture Expenditure, and Foursquare Features Changes of London Boroughs in Later Years }
\end{figure*}

\section{Metrics}

\subsection{Network Metrics}
The network measures used are in-degree centrality, outdegree centrality, and average clustering coefficient. Indegree centrality represents how many in-flow transitions the nodes of an area receive from other areas. Out-degree centrality of an area represents how many transitions start from this area, but end in other areas. The local clustering coefficient captures the degree to which the neighbours of a given node are connected with each other. For a node $i$ with degree $k_i$, the local clustering coefficient (Watts and Strogatz 1998) is defined as:

\begin{equation}
C_i = \frac{L_i}{k_i(k_i-1)}
\end{equation}

where $L_i$ represents the number of edges between the $k_i$ neighbours of node $i$. Then, average clustering coefficient, which reflects the overall level of clustering in an area is measured by averaging the local clustering coefficients of all the nodes within it.

\subsection{Cultural Advantage Metrics}
\subsubsection{Cultural Expenditure Advantage}

To evaluate the priority of cultural expenditure for a local area than the city average, we define \textit{cultural expenditure advantage (\textit{CEA})}. The \textit{CEA} for area $i$ is represented as:

\begin{equation}
CEA_i = \frac{CE_i}{TE_i}\cdot \left ( \frac{\sum _{i\in I}CE_i}{\sum _{i\in I} TE_i} \right )^{-1}
\end{equation}

where $CE_i$ is the cultural expenditure of neighbourhood $i$; $TE_i$ is the total expenditure of i; and $I$ is the set of neighbourhoods in the city.

\subsubsection{Cultural Venue Advantage}
Similarly, a metric of \textit{cultural venue advantage (\textit{CVA})} is given to measure the extent to which an area provides more cultural venues than the city average. Here, cultural venue includes 58 categories of Foursquare culture related venues, such as museums, art galleries, theatres, libraries, and parks. The \textit{CVA} for neighbourhood $i$ can be defined as:

\begin{equation}
CVA_i = \frac{CV_i}{TV_i}\cdot \left ( \frac{\sum _{i\in I}CV_i}{\sum _{i\in I} TV_i} \right )^{-1}
\end{equation}

where $CV_i$ is the number of cultural venues in $i$ and $TV_i$ is the number of all venues in $I$.

\section{Evaluation \& Results}
\subsection{Socio-cultural Neighbourhood Groups}
In this subsection, we provide a general view at borough level to exam the observability of cultural investments’ effects in the geo-social data and investigate whether different socio-economic status and cultural expenditure in neighbourhoods would result in significantly different properties in the geo-social network graph. This study lays the foundation for deeper discussions on smaller areas. 

In Figure 1, we reveal the relationship between IMD and cultural advantage in 2010. Here, the colour bar presents IMD scores in 2010, where higher numbers indicate more deprived areas. As we can see from this plot, yellow bubbles cluster in the middle/lower part, suggesting more deprived boroughs spent relatively less on culture and showed average cultural venue advantage in the beginning. Then in Figure 2, we discuss how the local authorities spent their money in the next two years, and how network features changed accordingly. It can be observed that more deprived boroughs spent more money on culture and had larger number of venues and transitions created.

Overall, boroughs with different initial IMD scores and culture expenditure priorities show different network patterns. Based on this observation, we identify four groups of London wards outlined in Table 2 for further investigation. Here, wards are firstly grouped into more deprived and less deprived classes according to whether their deprivation level in 2010 is higher or lower than the city average. Then, they are further classified according to their cultural spending priority. Specifically, if the \textit{CEA} of a ward is more than 1, it is clustered into the more advantaged group; otherwise, it is put into the less advantaged group.

\begin{table}[t]
\footnotesize
	\centering
	\begin{tabular}{*{5}{c}}
		\hline
		  & \textbf{Group 1} & \textbf{Group 2} & \textbf{Group 3} & \textbf{Group4} \\
		\hline
	    \textbf{\makecell{Initial\\ IMD}} & \makecell{less\\deprived} & \makecell{more \\deprived}& \makecell{more \\ deprived} & \makecell{less \\ deprived}\\
		\hline
		\textbf{\textit{CEA}}& \makecell{more \\ advantaged} & \makecell{less \\ avantaged} & \makecell{more \\ advantaged} & \makecell{less\\advantaged}\\
		\hline
		\textbf{\textit{N}} & 160 & 192 & 88 & 114\\
		\hline
	\end{tabular}
	\caption{Groups of London Wards in ANOVA Analyses.}
\end{table}

\begin{figure*}
	\centering
	\includegraphics{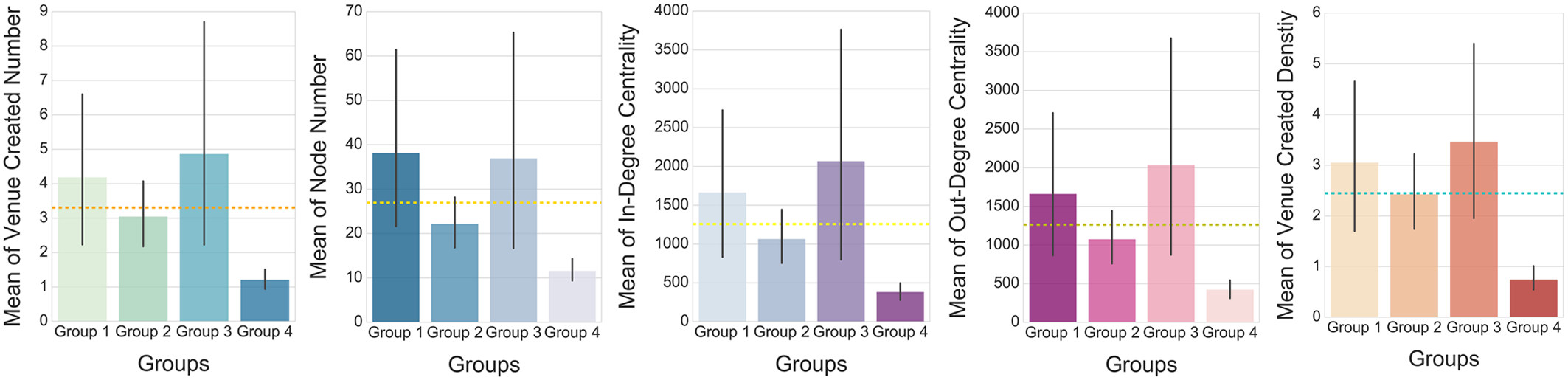}
	\caption{Means Plots for Five Variables with Statistically Significant Effects in Independent One-way ANOVA Analysis}
\end{figure*}

\begin{figure*}
	\centering
	\includegraphics{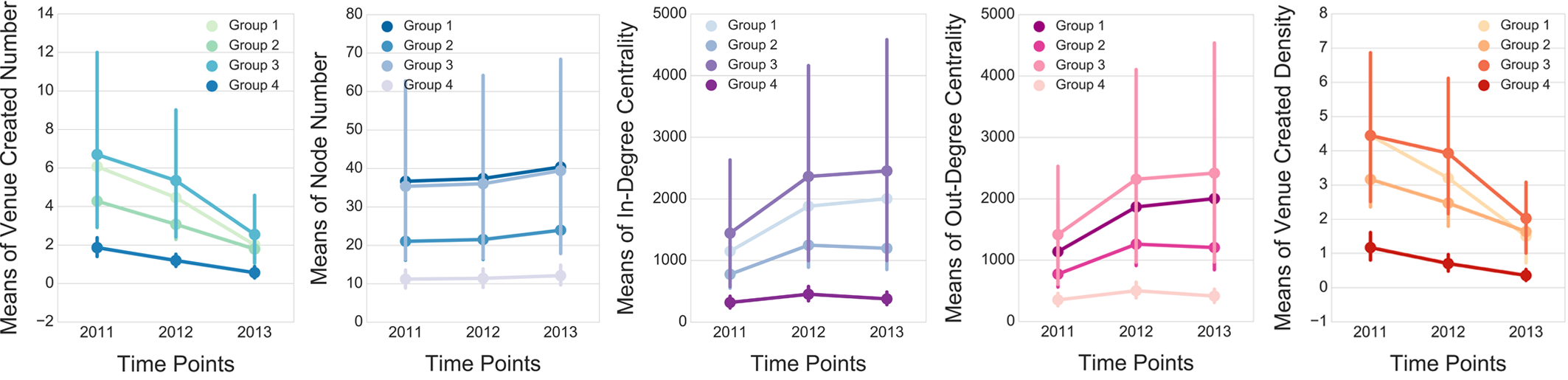}
	\caption{ Means Plots for Variables with Statistically Significant Effects in Factorial Repeated Measures ANOVA Analysis}
\end{figure*}

\subsection{Hypotheses Evaluation}
In the following hypotheses evaluation, two types of ANOVA analyses will be applied to get a deeper understanding
of the patterns we found at borough level, and test how they vary between ward groups and time periods.

\subsubsection{[H1] Network and Local Properties.}
To test [H1], we employ independent one-way ANOVA to examine whether statistically significant differences exist in terms of a set of network and local features between different ward groups.

From the output results, we can see that there are significant effects of groups on six features at the $p<.05$ level. The factor that distinguishes less and more deprived groups is average clustering coefficient with a statistically significant main effect of $F(3,550)=4.15, p=.006$. Furthermore, less deprived wards have higher means of average clustering coefficient. The other five features reveal significant difference between cultural advantaged and disadvantaged groups. Their means are plotted in Figure 3, from which we can find areas that gave higher priority to culture had larger venue created number, node number, in-degree centrality, out-degree centrality, and venue created density on average. Additionally, Group 3, that was more deprived in 2010, but invested relatively larger amount of money in culture later, performed best in most cases.

Through One-way ANOVA analysis, we find that urban areas with high cultural investment and deprivation level show significantly higher values in network properties than those areas with low cultural investment and deprivation level, which suggests that [H1] is true.

\subsubsection{[H2] Growth of Network and Local Properties.}
After discussing the overall differences between groups, we test whether groups experienced significant differences in growth with respect to network and local properties in this part. Technically, we examine whether there are statistically significant differences between different years and whether interaction effects exist between group and time point by using factorial repeated measures ANOVA. 

Figure 4 gives the means plots of the features that show significant effects. It can be found that Group 3 and Group 1 had dramatic advantages for all the three years. To consider groups separately, Group 3 had the highest values, while Group 4 had the lowest. In addition, significant interaction effects between group and year are found in three features: venue created density ($p=.008$), in-degree centrality ($p=.038$), and out-degree centrality ($p=.037$).

In this subsection, we detect how the growth of network and local properties varied between groups and confirm [H2] by finding that areas with high cultural investment and deprivation level experience significant growth with respect to network and local properties from areas with lower cultural investment and deprivation level.

\section{Discussion \& Conclusions}

In this work, we have investigated the socio-economic impact of cultural expenditure on London neighbourhoods, visible through the lens of location-based mobile data. We propose an innovative approach to giving insights on underlying relationships between socio-economic status, cultural investment and geo-social network properties. By applying new cultural metrics and traditional network metrics to the geo-social graph of transitions between venues on Foursquare, we show that areas with high cultural investment and deprivation experience significant growth. It
will have significant implications for location-based mobile systems, local governments and policymakers alike.

\section{Acknowledgement}

We would like to thank Foursquare for supporting this research by providing the dataset employed in the analysis.

\subsection{References} 

\smallskip \noindent \\
Eagle, N., Macy, M. and Claxton, R. 2010. Network diversity and economic development, \textit{science} 328(5981): 1029-1031.
\smallskip \noindent \\
Gonzlez, S. 2011. Bilbao and Barcelona ‘in motion’. How urban regeneration ‘models’ travel and mutate in the global flows of poli-cy tourism,  \textit{Urban Studies} 48(7): 1397-1418.
\smallskip \noindent \\
Hristova, D. Williams, M. J. Musolesi, M. Panzarasa, P. and Masco-lo, C. 2016. Measuring Urban Social Diversity Using Interconnected Geo-Social Networks. In  \textit{Proceedings of the 25th International Conference on World Wide Web}, 21-30.
\smallskip \noindent \\
Quercia, D. and Saez, D. 2014. Mining urban deprivation from foursquare: Implicit crowdsourcing of city land use,  \textit{IEEE Pervasive Computing} 13(2): 30-36.
\smallskip \noindent \\
Quercia, D., Seaghdha, D. and Crowcroft, J. 2012. Talk of the city: Our tweets, our community happiness. In  \textit{Proceedings of the 6th international AAAI Conference on weblogs and social media}. 
\smallskip \noindent \\
Smith, C. Mashhadi, A. and Capra, L. 2013. Ubiquitous sensing for mapping poverty in developing countries, Paper submitted to  \textit{the Orange D4D Challenge}.
\smallskip \noindent \\
Venerandi, A., Quattrone, G., Capra, L., Quercia, D. and Saez-Trumper, D. 2015. Measuring urban deprivation from user generat-ed content, in  \textit{Proceedings of the 18th ACM Conference on Comput-er Supported Cooperative Work \& Social Computing}, 254-264.
\smallskip \noindent \\
Watts, D. J. and Strogatz, S. H. 1998. Collective dynamics of ‘small-world’ networks,  \textit{nature} 393(6684): 440-442.

\end{document}